\documentclass[12pt]{iopart}
\usepackage{iopams}
\usepackage[]{graphicx}
\usepackage{float}

\begin{document}

\title[]{
Assessing structural bonding aspects of multiband superconductors 
through impurity-induced local lattice distortions:
a case study on ${\rm\bf{MgB_2}}$
        }

\author{Aleksandr Pishtshev$^1$ and Mihhail Klopov$^2$}
\address{$^1$ Institute of Physics, University of Tartu, Riia 142, 51014 Tartu, Estonia}
\ead{aleksandr.pishtshev@ut.ee}
\address{$^2$ Department of Physics, Faculty of Science, 
Tallinn University of Technology, Ehitajate 5, 19086 Tallinn, Estonia}
\ead{mihhail.klopov@ttu.ee}

\begin{abstract}
We report the results of an ab initio modeling of
substitutional impurities such as zinc, copper and zirconium ions
incorporated into the magnesium sublattice of $\rm{MgB_2}$.
These species are of particular interest because 
they demonstrate the lowest-observed-droop of the superconducting
temperature among most of the impurities of the same formal valences.
The goal of computational studies was to gain an atomistic understanding of how
the structural and bonding properties of the local ionic environment
are affected by a change in the host cation.
The simulations performed for the given set of substituents
indicate that at a low doping level,
the induced lattice distortions and additional forces are noticeable small.
At the same time, the electron redistributions around the impurities
have been found to differ drastically both from that around the host cation and
from each other.
Results of the first-principles calculations were used to compare and discuss,
in the context of the cation properties, specific changes in local charge structures
and in chemical bonding caused by the presence of the impurity.
Certain trends in substituent effects were found.
It was determined how the substitutional ion modifies
the initial charge distribution around the host cation position
and therefore, affects both the overall picture of the electronic states involved 
in the chemical bonding and the intrinsic charge-transfer channels.
Our studies demonstrate that
the robust structural stability of the boron sublattice of $\rm{MgB_2}$,
which preserves the nearest-neighbor environment around the $\rm{B}$ atoms practically
unchanged from an impurity incorporation into the magnesium site,
has important functional significance -- it supports a much weaker dependence 
of the superconducting temperature on impurity content.
The present work gives insight into structural and electronic features
of the impurity caused-processes which contribute to unique material properties
of various $\rm{MgB_2}$-solid solutions.
\end{abstract}

\maketitle

\section{Introduction}
In a theoretical analysis of impurity signatures, principally important
is the fact that the equilibrium between the (nearest neighbors) lattice atoms, and
the implanted impurity atom represents the structural balance between local geometry
and chemical bonding, which depends strongly on the interatomic distances.
In order to explore the dynamical nature of substituent effects on small length scales
one can employ an approach based on a detailed atomistic view:
it can be straightforwardly implemented within ab-initio numerical simulations,
which are able to calculate the relative ionic displacements and the charge density
redistribution accompanying the effect of the substitution.
The results of calculations will then allow us both to capture the final binding
geometry around the impurity site and to answer the fundamental question
as to what extent the impurity atom is capable of adapting itself to the local
environment and the existing chemical bonds.

In this paper, we apply such integrative approach to investigate the effects of
substitutions for several $\rm{3d}$
transition-metal impurities such as divalent $\rm Cu(II)$, $\rm Zn(II)$ and tetravalent
$\rm Zr(IV)$ in magnesium diboride $\rm{MgB_2}$.
We consider $\rm{MgB_2}$ containing substitutions of $\rm{Mg}$ as a doped model system
for which we perform a series of the relevant numerical simulations. 
There are two reasons that make this system attractive for detailed study.
First, while different properties of magnesium diboride doped by a variety of impurity
atoms have been extensively studied in the past decade, the investigation of
impurities in this material is still a topical subject,
raising a number of actual issues that are worthy of special attention (for details
see~\cite{Buz,Rav,Mazin1,Ivan,Cava,Singh2,Bern,Castro,Kortus2,Kuzm,WLi,Ojha},
and the references cited therein).
The other reason is that knowledge on how the arrangement of a single impurity ion
influences on a local structural order may have useful predictive value 
in the rational design of magnesium diboride solid solutions.
One of those issues, which is especially important to all aspects of local geometry
and stability, is the influence of substitutions on a principal atomic
configuration and the bonding properties of the cation environment.
However, in many cases, information on the bonding situation around a
substitutional impurity turns out to be limited, and there remain several questions 
about binding of metal elements incorporated into the place of 
the host $\rm{Mg(II)}$ cation.
Understanding the latter is very essential because the response of the system
on doping is not clear {\it a priori}. For example, 
it was experimentally established that the suppression of the superconducting
properties proceeds much slower in the $\rm Zn$-, $\rm Cu$- and $\rm Zr$-doped systems,
as compared to compounds that contain other dopant ions of the same formal valences
(details can be found in~\cite{Buz,Ivan,Singh2,Kortus2,WLi,Singh1}).
In table 1, using the data available from the literature, we have summarized 
some examples related to this observation.
\begin{table}
\caption{\label{table1}Experimental lowering of the $T_c$ value
($\Delta{T_c}$) in the $\rm M_{x}Mg_{1-x}B_{2}$ compositions
at substitution levels ($x$) in the range $0.03 < x < 0.10$.}
\footnotesize\rm
\lineup
\begin{tabular}{@{}rcccccccc}
\br
$M$ & ${\rm Zn}$ & ${\rm Cu}$ & ${\rm Zr}$ & ${\rm Ca}$ & ${\rm Co}$ & ${\rm Fe}$ 
 & ${\rm Ni}$ & ${\rm Pb}$ \\
\mr
$x:$ $\Delta{T_c}$ 
 & $0.03: -0.2$$^{\rm a}$ & $0.03: 0.5$$^{\rm c}$ & $0.05: 0.5$$^{\rm e}$ & 
  $0.03: 0.5$$^{\rm g}$ & $0.03: 2.8$$^{\rm h}$ & $0.03: 5.1$$^{\rm i}$ 
& $0.03: 2.9$$^{\rm i}$& $0.03: 1.5$$^{\rm j}$ \\
\mr
 &$0.05: 0.5$$^{\rm b}$ & $0.05: 0.2$$^{\rm d}$& $0.05: 0.7$$^{\rm f}$ &
  $0.05: 1.2$$^{\rm g}$ & $0.03: 2.0$$^{\rm i}$ & $0.05: 8.5$$^{\rm i}$ 
                           & $0.05: 4.8$$^{\rm i}$ & $0.05: 1.8$$^{\rm j}$ \\
\mr
&$0.10: 0.2$$^{\rm b}$ & & $0.10: 1.1$$^{\rm f}$ &
 $0.07: 1.8$$^{\rm g}$ & $0.05: 3.3$$^{\rm i}$ & & & $0.07: 2.2$$^{\rm j}$ \\
\br
\end{tabular}
\\
$^{\rm a}$\cite{Mor}; $^{\rm b}$\cite{Kaz};
$^{\rm c}$\cite{Tam};
$^{\rm d}$\cite{Kir};
$^{\rm e}$\cite{Kal};
$^{\rm f}$\cite{Nar};
$^{\rm g}$\cite{Sun};
$^{\rm h}$\cite{Shi};
$^{\rm i}$\cite{WLi};
$^{\rm j}$\cite{AM}
\end{table}

From the rate of decrease of the superconducting temperature $T_c$ with chemical
substitution of the original $\rm Mg$ with metal elements in $\rm{MgB_2}$
one can expect that not all the impurity properties are matched well 
with those of $\rm{Mg}$.
Thus, the given example of table 1 shows that conceptually
$\rm Zn$-, $\rm Cu$- and $\rm Zr$-containing solutions stand among
other doped systems as a particular example. 
This feature prompted us to examine in more detail the role of these substitutional
ions and ion-specific effects in the context of impurity-induced changes in local
structure, dynamics and chemical bonding.
Therefore, the aim of the present paper is by employing ab initio calculations
to study variations of certain lattice and charge characteristics, such as
lattice distortions, interatomic forces, charge densities and bonding features,
caused by compositional variations in $\rm{MgB_2}$.
By comparative analysis of features of the impurity atom with those related
to the host one
we pay attention in our work to what is happening within an ionic environment
in the elementary cell when an intensity of interionic and electronic correlations
is affected by the replacements of a magnesium cation.
\section{Computational details and impurity simulation}
According to the literature data (e.g.,~\cite{Buz,Ivan,Mazin2}),
a crystalline structure of the bulk $\rm{MgB_2}$
consists of alternating honeycomb layers of $\rm{B}$ interleaved with closed packed
layers of $\rm{Mg}$. Each $\rm{Mg}$ atom is located at the center 
of a hexagon formed by boron; by donating two valence electrons
it becomes a strongly charged ion with a charge very close to $+2$.
Consensus on the nature of the superconductivity in $\rm{MgB_2}$ is based on
a generic multiband model which employs the intra- and interband pair-scattering
mechanisms.
The presence of the interband channel connecting 
the $\sigma$ and the $\pi$ electronic bands~\cite{NN1,NN2,BHB}
gives rise to a unique property of $\rm{MgB_2}$ - two-gap superconductivity,
which is characterized by two gaps of different sizes 
that are simultaneously closed
at the same temperature $T_c$ of the superconducting transition.

The first-principles modeling of the effects of 
$\rm Mg(II)$ ion substitutions in the $\rm M_{x}Mg_{1-x}B_{2}$ compositions,
where $\rm M$ stands for a transition metal cation such as 
$\rm Zn$, $\rm Cu$ or $\rm Zr$, was performed using 
the Vienna Ab-initio Simulation Package
(VASP)~\cite{Kresse1,Kresse2,Kresse3,Kresse4} in conjunction with
a plane-wave-basis code and the projector-augmented wave (PAW)
method~\cite{Blochl,Kresse5}. Exchange-correlation effects were described through 
the Perdew-Burke-Ernzerhof (PBE) framework~\cite{Perdew}.
For geometry  optimization, one of the host atoms was replaced by 
one $\rm M$ element in a $81$-atoms supercell ($3\times3\times3$).
Such arrangement shown in figure 1 corresponds to $x=0.037$ 
in terms of the formal stoichiometry of the doped system.
The lattice distortion caused by the substitution has the following features:
(1) it is totally symmetric on the cation site, and (2) it involves
displacements of $\rm 12$ nearest-neighbor boron atoms from
regular positions. For a central boron atom, these displacements lead in turn
to local symmetry breaking.
Full relaxation of atomic positions was made with periodic boundary conditions and
theoretical (structurally relaxed) bulk lattice parameters.
The convergence criteria for total energy and residual forces
calculations were set to be within $1\times10^{-3}$ $\rm eV/atom$
and $1\times10^{-3}$ eV/\AA, respectively.
A cut-off energy of $700$ eV and a Monkhorst-Pack $k$-point grid~\cite{Monk}
of $\rm{5\times5\times5}$ has been employed.
This computational scheme was successfully tested by a comparison of
the calculated values of the lattice constants,
$a_{calc}=3.069$ \AA; $c_{calc}=3.504$ \AA; $c_{calc}/a_{calc}=1.142$,
with the experimental data~\cite{Mazin2},
$a=3.085$ \AA; $c=3.523$ \AA; $c/a=1.142$ (the deviations
are $0.52\%$ and $0.53\%$, respectively).

Changes in chemical bonding upon incorporation of a dopant atom into 
a $\rm Mg$-site was studied in terms of charge-density distributions.
In order to characterize variations of the partial charges at the magnesium
position upon substitution and to assign a specific meaning of ionic charges
in the analysis, values of the Bader effective charges~\cite{Bader} have been
evaluated. Their calculations were carried out by using software~\cite{Henk},
which utilizes a grid-based Bader analysis algorithm~\cite{Tang}.
The contribution of local many-body polarization effects was analyzed 
in terms of the Born effective charges, values of which have been 
obtained by using the special extended method adopted at VASP.
\section{Results}
In tables 2 to 5, we present the main results which characterize
the influence of the ${\rm Zn}$, ${\rm Cu}$ and ${\rm Zr}$ substitutional
impurities on displacements of the boron atoms,
the electronic charge redistribution
around a cation, and the local properties of chemical bonding in ${\rm MgB_2}$.
These results and several consequences of magnesium replacement
are discussed below.
The main conclusions are summarized at the end of the paper.
\subsection{Characterization of local distortions}
\begin{table}
\caption{\label{table2}Calculated lattice distortions
and changes of interatomic forces in the boron layer caused
by substitutional impurities in the magnesium diboride$^{\rm a}$.}
\footnotesize\rm
\lineup
\begin{indented}\item[]
\begin{tabular}{@{}lccccc}
\br
&&&& \makebox[0cm][r]{force components} \\
Host/Imp. & B out& B in & along $c$-axis & in $a$-$b$ plane \\
cation    & (\AA)& (\AA)& (eV/\AA) & (eV/\AA) \\
\mr
$\rm{Mg}$ & $0$ & $0$ & $0$ & $0$ \\
\mr
$\rm{Zn}$ & $+0.0045$ & $-0.0054$ & $+0.020$ & $-0.089$ \\
\mr
$\rm{Cu}$ & $-0.0002$ & $-0.0128$ & $-0.014$ & $-0.234$ \\
\mr
$\rm{Zr}$ & $-0.0063$ & $+0.0159$ & $-0.013$ & $+0.260$ \\
\br
\end{tabular}
\\
$^{\rm a}$Denoted: 'B out/B in' -- boron out-of/in-plane displacement
as shown in figure 1; the minus sign indicates inward directions.
\end{indented}
\end{table}

We pay attention to the data of table 2 as they reveal certain trends.
First, no significant changes are seen -- local lattice distortions, and the induced
changes in transverse ionic strengths caused by cationic substitution appear
to be notably small. Particularly, this observation
relates to cases with $\rm{Cu}$ and $\rm Zn$.
This implies that the relevant bond lengths are preserved, 
and the boron layer remains planar.
Further, it is also seen from table 2 that longitudinal (in-plane) stresses
are more strongly affected by the impurities, though the induced changes
remain relatively small.

The overall comparison of the data of table 1 with the simulation results of table 2
reveals a certain correlation between the variation in the magnitude of
the $T_c$ suppression rate (${\Delta}T_{c}/{\Delta}x$) and the doping induced changes
in the boron sublattice. This can be emphasized by observing the following difference:
While in the majority of the doped solutions, the presence of several percents of
divalent impurity metal atoms provides already a considerable suppression of $T_c$,
there are substitutions such as $\rm Zn$, $\rm Cu$, or even tetravalent $\rm Zr$
low concentrations of which slightly influence both the superconducting order
and the specific positions of the boron atoms.
The reason for this discrepancy seems to be related primarily to
the strong $\rm 2D$ $\rm B$-$\rm B$ linkage dominating in the boron layer~\cite{Choi}
which, as we have shown above, tends to keep the existing structural stability
upon a change in the host cation.
Considering $\rm Mg$ as a reference cation, another interesting result is that
one may say according to~\cite{Mazin1,Mitr} that the incorporation of
just $\rm Zn$, $\rm Cu$ or $\rm Zr$ into the $\rm Mg$-site 
has little effect on actual interband pairing channels.
Correspondingly, in these cases, the protection function of the covalent 
boron linkage cannot be markedly influenced (at low doping levels)
what may provide a much weaker dependence of 
the superconducting temperature $T_c$ on impurity content.
\subsection{Characterization of dopant local bonding}
In order to get further insight into why the substitutional $\rm Zn$, $\rm Cu$ and 
$\rm Zr$ cause only minor changes in a planar geometry of the covalent boron layer,
we undertook an investigation of how local properties of chemical
bonding are affected by the overall electronic structure and elemental characteristics
of each of the considered impurity ions.
For characterization of impurity-induced changes in chemical bonding via 
the differences of the electron states between the host and foreign cations,
we performed an extensive comparative analysis of the relevant 
charge partitioning schemes.
Figure 2 shows charge density maps calculated for $\rm{MgB_2}$ and all three doped
cases; the maps are projected onto the $\rm (100)$ plane and contain
the central $\rm{Mg}$ or $\rm{M=Zn,Cu,Zr}$ cations, and surrounding $\rm{B}$ atoms.
Figure 3 is instructive in another way: for each cation considered, it shows 
the line charge density distributions along the $\rm Mg$-$\rm B$ and $\rm M$-$\rm B$
directions, respectively.
Computed values of the Bader effective charges ($Q_{B}$) presented in table 3
reflect the topology features of the underlying electron-density distribution
emerging as a direct consequence of the relevant arrangements.
\begin{table}
\caption{\label{table3}
Some charge characteristics related to the host cation and its substituents
in the magnesium diboride.
The Pauling electronegativities are taken from~\cite{EnvChem}.
The quantities ${\rm Q_{B}(M)}$ denote Bader effective charges calculated
from electronic densities (in units of $\left|e\right|$).}
\footnotesize\rm
\lineup
\begin{indented} \item[]
\begin{tabular}{@{}lcccc}
\br
 & ${\rm Mg}$ & ${\rm Zn}$ & ${\rm Cu}$ & ${\rm Zr}$ \\
\mr
Electronegativity difference & & & & \\ 
values with respect to boron  & $0.73$ & $0.39$ & $0.14$ & $0.71$ \\
\mr
${\rm Q_{B}(M)}$ & $+2.0$ & $+0.69$ & $+0.42$ & $+2.68$$^{\rm a}$ \\
\br
\end{tabular}
\\
$^{\rm a}$A comparison shows that this value appears to be equal to that of
$+2.67$ obtained from the self-consistent LAPW calculations for the bulk 
zirconium diboride $\rm{ZrB_2}$~\cite{Swit}.
\end{indented}
\end{table}
Comparison of figures 2, 3 and the data of table 3 
for the reference case of the $\rm Mg(II)$ ion reproduces
the well-established result~\cite{Rav,Mazin2,An,Kortus1,Bel,DeLaMora}
that $\rm Mg$ is an ionically-bonded cation (with an admix of some weak covalency)
in a stable divalent state.

Analysis of cases where $\rm Zn$ or $\rm Cu$ substitutes for $\rm Mg$
revealed three significant differences.
First, as depicted in table 3, the Bader effective charges
are quite different from the formal chemical charge $\rm{+2}$. 
Two other differences are very clearly delineated on figures 2 and 3;
common to both ions they are: a depletion of the electron charge along
the metal-boron bonding direction, and a large degree of electron charge
accumulation of fairly good spherical form around the nuclei.
Correspondingly, the physical picture of the outer electronic structure of 
$\rm Zn$ and $\rm Cu$ incorporated into magnesium diboride can be drawn as follows:
(i) the $\rm 3d$ states appear to be completely filled (what provides
increased electron density at the substituent, causes the observed 
spherical charge distribution around it, and induces the smaller effective
charge transfer, in comparison with magnesium),
(ii) zinc and copper cations acquire fractional quantities of 
the $\rm 4s$ electron populations, about $\rm 1.31e$ and $\rm 0.58e$, respectively,
(iii) the partially filled $\rm 4s$ states constitute the resulting outermost orbitals,
and (iv) the $\rm d$-electron covalency effects are extremely low 
because the filled $\rm 3d$ levels of the substituent are shielded 
by its $\rm 4s$ states.
Remarkably, smallness of the effective charges relative to the nominal chemical
valence $\rm +2$ is an indication of the fact that the substitutional cations 
have retracted the certain amount of the delocalized electrons that participate 
in metallic bonds.
The reason why just the electrons shared by the boron atoms are excepted is that 
the significant changes in an equilibrium planar geometry of the covalent boron
layer are absent.
Furthermore, of note is another observation from table 3 that 
the local loss of charge exchange activity is directly connected with
the drop in the electronegativity difference with
respect to boron in the series of elements $\rm Mg$, $\rm Zn$, $\rm Cu$.

It may be concluded thus that both substitutions give rise to
a negative tendency which manifests itself (i) in increasing the electron
density around outer orbitals of the $\rm Zn$ or $\rm Cu$ cations,
and (ii) in decreasing the positive charge at the cation center up
to the values of ${\rm Q_{B}(Zn)}=+0.69$ and ${\rm Q_{B}(Cu)}=+0.42$, respectively.
This leads to partial neutralization of the overall charge on the substituents, 
makes the initial charge transfer channels much weaker and therefore 
causes the significant reduction of the polar contribution into
the cationic nature of the substitutional ion.
As a result, the initial, predominantly ionic character of the local environment 
is degraded.
The variances of the Bader effective charges from the nominal $\rm{+2}$ may therefore 
be regarded as a degree of $\rm Zn$ or $\rm Cu$ ``underbonding'' that should have
a certain effect on physical properties of $\rm{MgB_2}$ doped by these elements.
For example, since the interband channel is unaltered (because of
minor changes in a planar geometry of the covalent boron layer),
the experimentally observed smallness of suppression of
the superconducting temperature is perfectly understandable via influence of 
the above-mentioned retraction of ``previously expelled electrons''
on the density of electronic states at the Fermi level.
This result correlates well with the conclusion made in~\cite{Moni1} for a series of
$\rm Al$-$\rm Li$ codoped $\rm{MgB_2}$ systems that the superconductivity is much
more affected by the relevant lattice distortion induced by impurity ion substitution
than by the band filling.

The trend towards electron density accumulation and a weakening of charge
transfer channels breaks down for the case of the substitutional $\rm{Zr}$
in the structure of $\rm{MgB_2}$. 
Although $\rm{Zr}$ possesses almost the same electronegativity difference
as $\rm{Mg}$ , the calculated value of the Bader effective charge,
${\rm Q_{B}(Zr)}=+2.68$, is considerably different from the formal
valence $4+$. From chemical point of view, this implies that 
the zirconium cation in the $\rm{MgB_2}$ lattice
tends towards a lower oxidation state\footnote{This point suggests 
the presence of some mechanism of valence reduction upon substitution,
as can be compared with $\rm{4d^{2}}$ configurations of low-valent $\rm{Zr(II)}$
in complex sheet and cluster structures~\cite{Cotton}.}.
In contrast to zinc and copper ions, $\rm{Zr}$ attempts to expand definite amounts
of charge density from itself toward an interstitial area.
Analysis of figures 2 and 3 strongly suggests the moderate enhancement 
of the positive charge on the cation which takes place due to delocalization
of its $\rm 4d$ electrons caused by partial donation of $\rm 0.68$ additional
electrons.
Such a delivery of a partial charge (along the metal--non-metal charge-transfer
configuration) is well correlated with a chemical readiness of 
the zirconium cation to give up the valence $\rm{4d}$ electrons
in order to move back to the predominant oxidation state of $\rm{4+}$
and to activate the covalent bonding with neighboring atoms.
Consequently, as compared with the parent compound $\rm{MgB_2}$, 
an ionization increase $\rm{Mg^{2+}}$ $\rightarrow$ $\rm{Zr^{2.68+}}$ 
on a substitutional cation site and the presence of appreciable covalency 
in the same region
may give rise to a tendency towards dielectrization of the system.
This observation is confirmed by first-principles 
studies of the bulk compound $\rm{ZrB_2}$~\cite{DeLaMora,Shein1,Rosner},
which show the presence the strong hybridization between 
the $\rm{Zr}$ $\rm 4d$ and the $\rm{B}$ $\rm 2p$ states as well as
drastic fall of the density of electronic states at the Fermi level.

Thus, we can conclude that electronic charge distribution at the $\rm Mg$ position
in the bulk magnesium diboride is very sensitive to the cation replacement.
Firstly, the substitution for an impurity ion of the same nominal charge $\rm +2$
leads to a substantial charge redistribution around the cation.
Secondly, the change in the chemical bonding character 
depends on the type of metal ion and its electronegativity versus boron,
so that among substitutional ions considered two different trends can be distinguished.
All this underlines in the light of the bonding situation
that ion substitution not only always affects the existing chemical bonds
but also that the resulting charge rearrangement is exhibited quite differently 
for the cations considered in the present work.
Moreover, in the context of superconductivity, these features can seriously limit
the targeting efficiency of a number of potential substituents 
or even make them unfavorable towards the enhancement of superconducting properties.
\subsection{Pressure effects}
To complete the results obtained for impurity-induced changes in the cation charge
states, it should be interesting to study how these changes proceed under pressure.
For this purpose, the $\rm Zr$- and $\rm Zn$-doped case were considered.
We employed the same set of calculations as used throughout this work
and fully relaxed a doped system.
The resulting pictures of the charge density maps are shown in figure 4.
The comparison of the charge distributions with those shown in figure 2 reveals 
no significant changes between them.
The most remarkable difference is a slight increase of the Bader
effective charge of the $\rm Zr$ cation -- from $\rm +2.68$ at zero pressure
to $\rm +2.84$ under pressure of $\rm 16.5$ GPa (what evidently indicates 
the above-mentioned oxidation tendency to the tetravalent state).
This implies that the charge state of $\rm Zr$ remains relatively stable 
with compression until at least $\rm 16.5$ GPa.

The $\rm Zn$ substitutional cation also demonstrates the similar tendency to maintain
its charge state under pressure.
For example, the calculated change in the Bader effective charge
under compression of $\rm 14.7$ GPa, 
$\rm {\Delta}Q_{B}(Zn)=+0.04$, shows that due to contraction of the filled 
$\rm 3d$ levels, valence electrons do not favor extra oxidation.
As a result, the overall charge configuration
of $\rm Zn$ in magnesium diboride is essentially unaffected by pressure.
\subsection{The Born effective charges}
Another possible microscopic reason of lattice distortions stems from the fact
well-known in the theory of local phase transitions that distortion of 
local geometry around an impurity may be caused by sufficient strength of
the purely dynamical part of the charge transfer (e.g.,~\cite{NN5A,NN5B}).
Generally speaking, significance of dynamical charge transfer processes 
generated by vibronic mixing of the proper electronic states is one of 
most important characteristic features of polar crystals~\cite{NN3,NN4,BH1,AP}. 
On the atomic level, this feature is associated with many-body polarization effects
that contribute to ionic dipole polarization, and it reflects a mixed 
ionic(polar)-covalent character of the relevant chemical bonds~\cite{AP}.
Since the latter can be interpreted in terms of
the Born dynamic effective charges with respect to some reference (nominal) ionic
value (e.g.,~\cite{Lee,AP}), the calculations of the principal values of
the Born dynamic effective charge tensor $\rm Z^{*}$ can be employed as
a test of how electronic effects of the many-body nature translate
the polar contribution into ionic chemical bonding.
It is also of note that although the idea of calculation of the Born dynamic
effective charges for a conducting material may seem senseless at first glance,
it can be, due to logic behind the ionic bonds and charge transfer, 
reasonably useful for a mixed bonded solid, indicating how specific pieces of 
a composite picture of chemical bonding may be distinguished.

In table 4, we have presented the converged values of the Born effective charge
for the magnesium cation in $\rm{MgB_2}$ (due to metallic properties of $\rm{MgB_2}$, 
we focused on dynamic charges only for cations and not for boron atoms).
It may be assumed that the numerical convergence to stable values
of table 4 is guaranteed by the nearly
ionic character of the $\rm Mg$ sublattice which, in the spirit of layered 
superconductivity approach~\cite{Kresin,Var}, can be figuratively imagined as a stack of 
the intrinsic insulating layers.
Calculations have also predicted principal values of 
the electronic dielectric constant for the bulk magnesium diboride:
$\rm \epsilon^{xx}_{\infty}=\epsilon^{yy}_{\infty}=22.45$,
$\rm \epsilon^{zz}_{\infty}=5.77$.
The theoretical value $\rm \epsilon_{\infty}=16.9$, evaluated as an average
$\rm \epsilon_{\infty}=(1/3)\Sigma_{i}\epsilon^{ii}_{\infty}$,
is quite comparable with the experimental estimate of $\rm 11.9$~\cite{Castro}
thus proving the consistency of the present analysis.
\begin{table}
\caption{\label{table4} Principal values of the Born effective charge of $\rm Mg(II)$
in magnesium diboride (in units of $\left|e\right|$).
The diagonal form with ${\rm Z^{*}_{xx}(Mg)}={\rm Z^{*}_{yy}(Mg)}$
reflects hexagonal symmetry of the tensor $\rm Z^{*}(Mg)$.
The quantity ${\rm \bar{Z}^{*}(Mg)}$ represents the average over crystal axes: 
${\rm \bar{Z}^{*}(Mg)}=(1/3)({\rm Z^{*}_{xx}(Mg)}+{\rm Z^{*}_{yy}(Mg)}
+{\rm Z^{*}_{zz}(Mg)})$.}
\begin{indented}
\lineup
\item[]\begin{tabular}{@{}*{7}{cccc}}
\br
 ${\rm Z^{*}_{xx}(Mg)}$ & ${\rm Z^{*}_{yy}(Mg)}$ & ${\rm Z^{*}_{zz}(Mg)}$ &
 ${\rm \bar{Z}^{*}(Mg)}$ \\
\mr
 $ +1.87 $               & $ +1.87 $               & $ +2.31 $ & $ +2.02 $ \\
\br
\end{tabular}
\end{indented}
\end{table}

The tensorial nature of the Born effective charge provides subtle
information concerning differences in directions and strengths of dynamical
processes related to the ionic sites and caused by the polarization effects.
As seen by comparing the values of table 4, there exists
a certain anisotropy, as is indicated by the observation that 
in the directions perpendicular (the ${ab}$-plane) and parallel to the $c$-axis
the components of the dynamic charge differ markedly from the formal value of $2+$.
This suggests that, although on the average $\rm Mg$ remains quite well ionized 
in the full $N$-particle ensemble (${\rm \bar{Z}^{*}(Mg)}=+2.02$), 
(i) the attached effective potential, which gives rise to polarization force fields
acting in a local environment around magnesium, appears non-spherical, and 
(ii) due to deviation from the nominal ionic charge some contribution of the pure
dynamic covalent nature in the magnesium--boron bonding should be assumed.
The latter is entirely consistent 
with previous studies of dynamical covalency effects in $\rm{MgB_2}$~\cite{Deng}.

In order to analyze how the change in the host cation influences
the attached effective potential, we carried out the relevant calculations
of the Born effective charges
for the substitutional impurities we are considering here.
In contrast to the previous case of magnesium, the computational method
failed to give uniform results consistent with hexagonal symmetry.
In the physical sense, the lack of numerical consistency is not discouraging, and
should be considered as an additional confirmation of the drastic changes
of a charge state which arise around the host cation position due to its
replacement.
In other words, impact of the substitutions with such unavoidable outcomes as falling
of ionic character of chemical bonding, and the expansion of a metallic environment 
beyond the range of the boron layers~\cite{DeLaMora,Shein1} 
make it impossible to get from the method of calculations the target values
of dynamic charges for the foreign cations.
To obtain some approximate estimates of the Born effective charges,
we have constructed a smooth procedure based on averaging the obtained results over
the supercell, so that all the cations employed in the calculations are taken
into account. This can be done since overall charge balance is maintained.
Due to electroneutrality of the simulation supercell,
one can then expect that summation eliminates off-diagonal contributions,
suppresses discontinuities, random perturbations and other artificial symmetry effects 
and thus will return the weighted-average components of the Born effective charge tensor
in the original diagonal form.

In table 5, we have presented the numerical values of the dynamic charges
for $\rm{Zn}$, $\rm{Cu}$ and $\rm{Zr}$ substitutional cations as averages over
the supercell.
\begin{table}
\caption{\label{table5}
The averaged principal values of the Born effective charge tensor for
the foreign cation $\rm M=\rm{Zn}$,$\rm{Cu}$,$\rm{Zr}$ in magnesium diboride
(in units of $\left|e\right|$).
${\rm \bar{Z}^{*(avg)}(M)}=(1/3)({\rm Z^{*(avg)}_{xx}(M)}+{\rm Z^{*(avg)}_{yy}(M)}
+{\rm Z^{*(avg)}_{zz}(M)})$.}
\begin{indented}
\lineup
\item[]\begin{tabular}{@{}*{7}{lcccc}}
\br
 & ${\rm Z^{*(avg)}_{xx}(M)}$ & ${\rm Z^{*(avg)}_{yy}(M)}$ & ${\rm Z^{*(avg)}_{zz}(M)}$ &
 ${\rm \bar{Z}^{*(avg)}(M)}$ \\
\mr
$\rm{Zn}$ & $ +1.6617 $  & $ +1.6617 $ & $ +2.25 $ & $ +1.86 $ \\
\mr
$\rm{Cu}$ & $ +1.1378 $  & $ +1.1345 $ & $ +1.85 $ & $ +1.38 $ \\
\mr
$\rm{Zr}$ & $ +1.4018 $  & $ +1.4630 $ & $ +2.26 $ & $ +1.71 $ \\
\br
\end{tabular}
\end{indented}
\end{table}
It is seen that the remaining uncertainties on the averaged values of charges 
are really negligible, so that this type of approximation is quite reasonable.
Nevertheless, one additional aspect which should be discussed here is 
the sensitivity of the retrieval accuracy within such a calculation procedure.
First of all, we note that all off-diagonal contributions are exactly eliminated.
On the other hand, the required identity of the diagonal terms 
(${\rm Z^{*(avg)}_{xx}}={\rm Z^{*(avg)}_{yy}}$) is restored at different accuracy
levels: for zinc and copper, we get a more accurate retrieval than for zirconium.  
We believe that this result is not surprising because some discrepancy
in the case of zirconium suggests that due to the expansion of metallic bonds
into the cationic environment, the effect of charge fluctuations is not completely
eliminated after summation over the supercell.

The results listed in table 5 allow us to characterize the specific features of 
chemical bonds in terms of the Born effective charges as follows.
(i) A dynamic component in resulting chemical bonding becomes smaller upon cation
substitution (as compared with that of $\rm Mg$), and
(ii) this partial loss in initial dynamic covalency is caused by the relevant charge 
redistribution around the cation position.
At the same time, as seen from table 5, a rate of an anisotropy in the dynamics
of valence charge distribution remains relatively unchanged in the cases of zinc and
zirconium.
Moreover, of all three components just the $z$ component of the Born effective charge, 
which is particularly interesting for dynamical covalency effects in magnesium diboride,
tends to be close to the $c$-axis value of the $\rm Mg$ cation and therefore
appears, in comparison with in-plane $x$, $y$ components, to be more robust to
alteration of charge densities.
Following the same consideration as in~\cite{Ghosez}, this leads us to an understanding
regarding why lattice distortions induced by $\rm{Zn}$, $\rm{Cu}$ or $\rm{Zr}$ ions
are so small: the explanation is that the ionic strength changes induced by variations
of dynamical charge transfer
along the $c$-axis are not large enough to drive the relative boron displacements.
\section{Conclusions}
In order to understand the leading factors that govern adjustments of 
the lattice positions during accommodating the foreign cation in $\rm{MgB_2}$,
we investigated the induced deformations (lattice distortions)
in the local geometry of the boron sites.
The first-principles calculations were performed 
for three different substitutional cations such as
divalent $\rm Zn(II)$ and $\rm Cu(II)$, and tetravalent $\rm Zr(IV)$.
These species are of particular interest because 
they demonstrate the lowest-observed-droop of the superconducting
temperature among most of the impurities of the same formal valences.
Inspection of the relaxed lattice structures together with 
the results of the performed calculations allowed the following conclusions
to be drawn: (i) At a low doping level, the $\rm Zn$, $\rm Cu$ and 
$\rm Zr$ substitutions cause only negligible changes in the structural arrangement 
of the neighboring boron atoms.
(ii) Such an insignificant impact on the local geometry of the boron layer
is mostly attributed to two aspects.
The first is the observation that the existing in-plane $\rm B$-$\rm B$
covalent bonding is strong enough to keep effectively the boron atoms
non-shifted under substitution of magnesium with $\rm Zn$, $\rm Cu$ or $\rm Zr$.
The second is that dynamical covalency processes along the cation-boron bonding 
tend to be tolerant of changes in the host cation and, consequently,
cannot create a channel of instability in the boron layer.
Both aspects are important in view of the fact that
the structural balance between local geometry and
chemical bonding are related to the electron structure of the impurity and
to the bonding connections carried out by the impurity.
(iii) A role of small deformations of the local
geometry of the boron layer induced by the impurity ions
may be insufficient to influence alone on superconductivity of the host system.
(iv) Generally, our results emphasize that $\rm{MgB_2}$ is the perfect material having
a unique combination of an ideal planar layer structure, 
the predominantly ionic character of the cationic sublattice, 
and the high degree of chemical bonding stability.

In order to investigate the characteristic measures of impurity activity,
such as charge states and the impurity-induced changes in the chemical bonding,
we performed the comparison analysis of the underlying charge-transfer processes.
It has been discovered that the replacement of $\rm Mg(II)$ by
divalent $\rm Zn(II)$, $\rm Cu(II)$ or tetravalent $\rm Zr(IV)$ influences
substantially the character of the charge transfer,
so that the distributions of the charge around the impurities and 
degree of ionicity differ drastically both from that around the host cation and
also from each other.
It has been shown that the different cationic nature of the substituents 
dictates two entirely opposite directions along which the induced changes
in the local structure take place.
First, the change in the host cation leads to density contraction --
the substitutional $\rm Zn$ or $\rm Cu$ demonstrate
the considerable increase of the overall charge density due to 
retaining its valence electrons and the small amount of
charge transfer to the boron layers. Particularly, a larger accumulation of a charge
and the least charge transfer are observed for the copper ion.
It was argued that this process retracts the certain amount of the delocalized
electrons from metallic bonds what causes the observed decrease 
in the superconducting temperature values as the substitution level grows.
Second, the change in the host cation leads to charge delocalization -- no charge
accumulation is observed for $\rm Zr$.
The $\rm Zr$ incorporation into the $\rm Mg$ site of $\rm{MgB_2}$ tends 
(a) to preserve the initial charge transfer channels, 
and (b) by partial donation of $\rm 4d$ electrons to add some amounts of ionicity
and covalency into the local cation-anion coupling.
It was suggested that such increasing of covalency character does not favor 
superconducting properties of the host material.
The stability of charge states related to the impurity cations was tested 
against pressure: the pressure dependence of the charge transfer degree 
was found to be very weak.
The theoretically observed effective charges indicate that 
the substitutional ions do not possess their formal valence (oxidation state),
and are characterized by different levels of electron delocalization;
zirconium as an electron donor is indeed stronger than magnesium. 
As a consequence, by comparison with less positively charged zinc and copper, 
the behavior of zirconium remains much more tolerant (affine) to the usual ionic
distribution associated with the host magnesium cation.

Summing up, one can say that the induced local lattice distortions and changes
of the charge transfer degree may serve as structurally sensitive prognostic parameters 
for signaling how the superconductivity in $\rm{MgB_2}$-solid solutions
is affected by impurities, for clarifying the functional roles and efficiency
of such various factors as a geometry of the boron layer, the electronegativity
difference, ionic character of the local environment, relative affinities toward
magnesium and boron, etc.,
and on this basis for facilitating more precise theoretical predictions in
synthesis and characterization.
\ack{
The authors are grateful to Professor N. Kristoffel for suggesting the subject 
of this paper and for numerous helpful discussions and constructive comments.

This work was supported by the European Union through 
the European Regional Development Fund
(Centre of Excellence ``Mesosystems: Theory and Applications'', TK114).
It was also supported by the Estonian Science Foundation grant No 7296. }
%
%
%

\newpage

\section*{\large Figure Captions:}

\subsection*{{\bf Figure  1:}  
\rm 
A bulk $81$-atoms supercell modeling the $\rm M_{x}Mg_{1-x}B_{2}$ composition
for $x=0.037$. The $\rm{MgB_2}$ cell containing the impurity cation is sketched.
In-plane and out-of-plane boron displacements are indicated by arrows.}

\subsection*{{\bf Figure  2:} 
\rm Charge density  maps of (a) $\rm MgB_{2}$, (b) $\rm Zn_{0.037}Mg_{0.963}B_{2}$,
(c) $\rm Cu_{0.037}Mg_{0.963}B_{2}$ and (d) $\rm Zr_{0.037}Mg_{0.963}B_{2}$
projected onto the $\rm (100)$ plane. }

\subsection*{{\bf Figure  3:} 
\rm Comparison of the line charge density distributions
calculated along the cation-anion direction.
The dashed, dot-and-dash and solid curves represent $\rm{Zr}$-substituted,
$\rm{Cu}$-substituted and $\rm{Zn}$-substituted $\rm{MgB_2}$, respectively.
The inset shows the corresponding profile for the non-substituted $\rm{MgB_2}$. }

\subsection*{{\bf Figure  4:}
\rm Charge density maps of 
(a) $\rm{Zr}$-substituted and (b) $\rm{Zn}$-substituted $\rm{MgB_2}$
projected onto the $\rm (100)$ plane. Calculations were made for a pressure of
$16.5$ GPa and $14.7$ GPa, respectively.}

\newpage

\section*{\large Figures} 
\vspace{2cm}

\begin{figure}[H]
\includegraphics[width=0.8\textwidth]{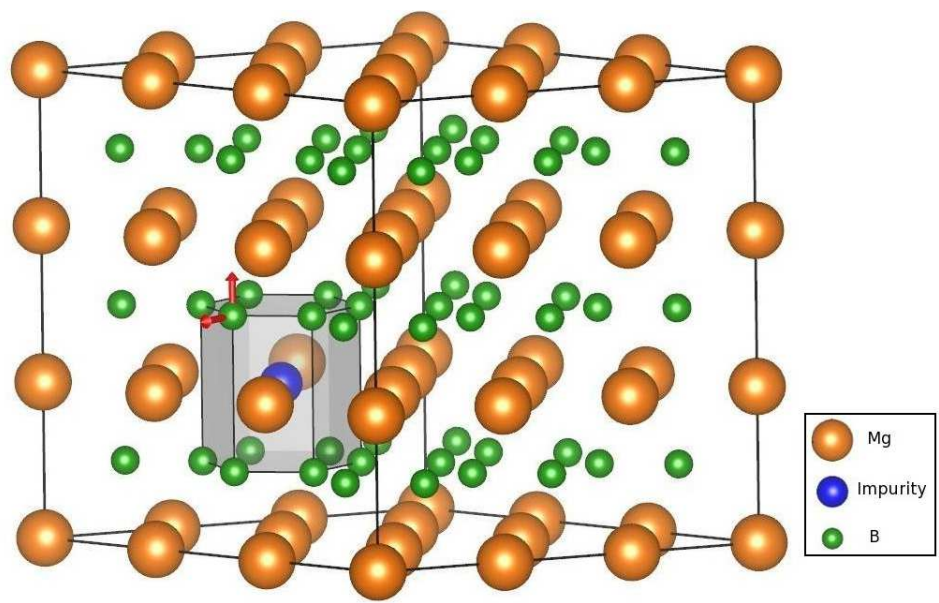}\hspace*{0em}
\vspace{2cm}
\caption{\rm 
A bulk $81$-atoms supercell modeling the $\rm M_{x}Mg_{1-x}B_{2}$ composition
for $x=0.037$. The $\rm{MgB_2}$ cell containing the impurity cation is sketched.
In-plane and out-of-plane boron displacements are indicated by arrows. }
\end{figure}

\newpage
\begin{figure}
\includegraphics[width=1.0\textwidth]{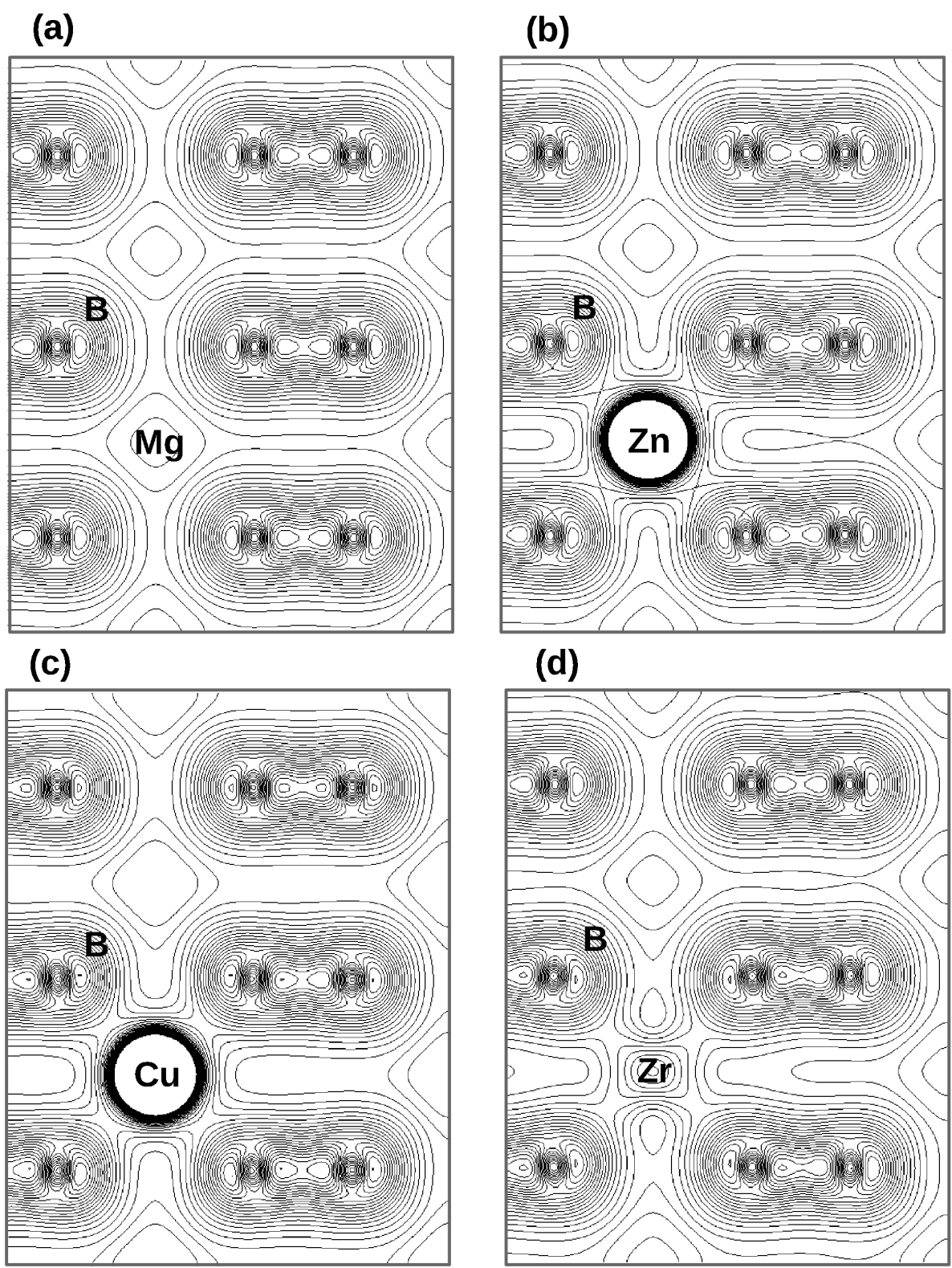}\hspace*{0em}
\vspace{2cm}
\caption{Charge density  maps of (a) $\rm MgB_{2}$, (b) $\rm Zn_{0.037}Mg_{0.963}B_{2}$,
(c) $\rm Cu_{0.037}Mg_{0.963}B_{2}$ and (d) $\rm Zr_{0.037}Mg_{0.963}B_{2}$
projected onto the $\rm (100)$ plane.}
\end{figure}

\newpage
\begin{figure}
\includegraphics[width=1.0\textwidth]{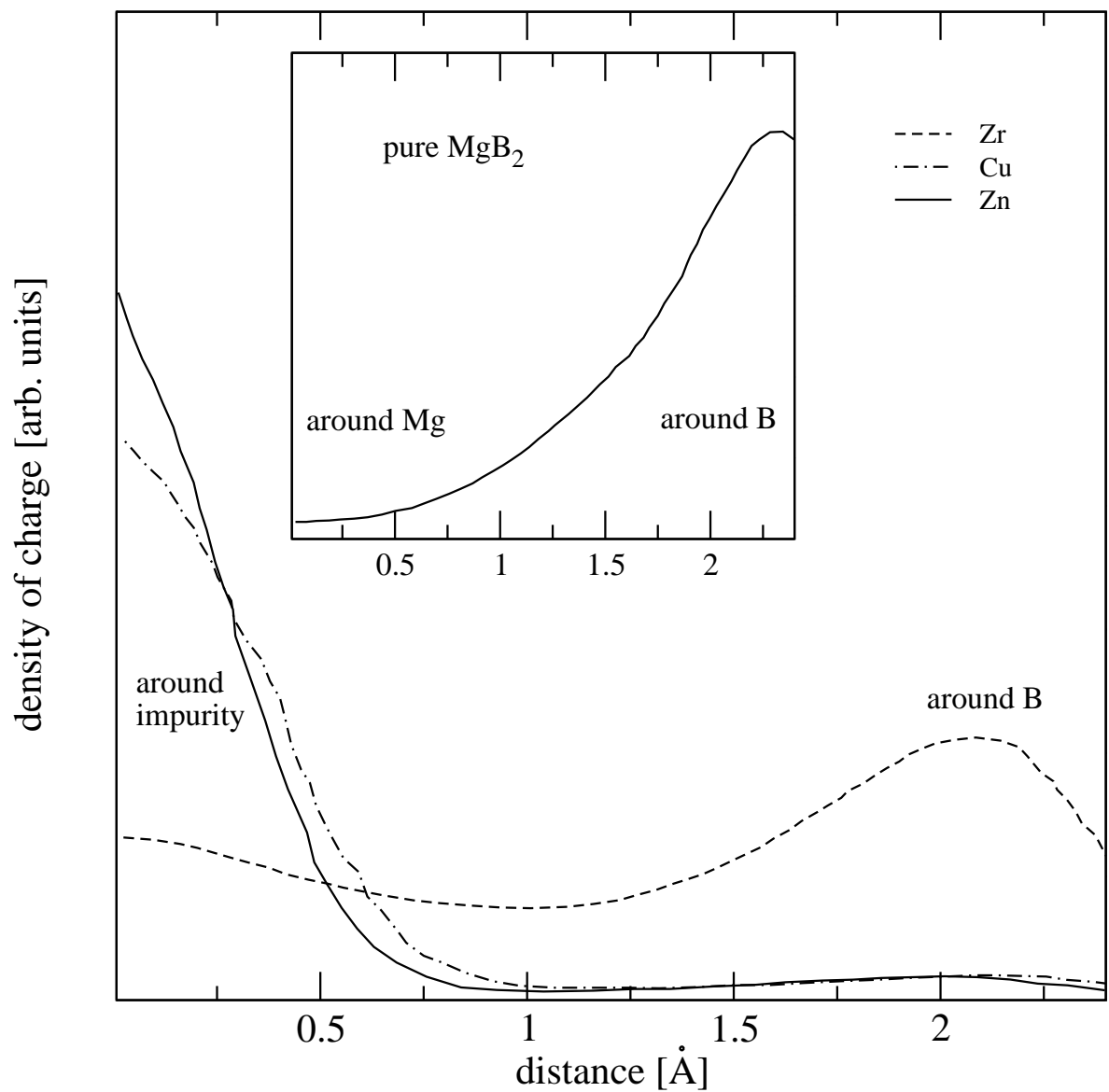}\hspace*{0em}
\vspace{1.5cm}
\caption{Comparison of the line charge density distributions
calculated along the cation-anion direction.
The dashed, dot-and-dash and solid curves represent $\rm{Zr}$-substituted,
$\rm{Cu}$-substituted and $\rm{Zn}$-substituted $\rm{MgB_2}$, respectively.
The inset shows the corresponding profile for the non-substituted $\rm{MgB_2}$.}
\end{figure}

\newpage
\begin{figure}
\vspace{2cm}
\includegraphics[width=1.0\textwidth]{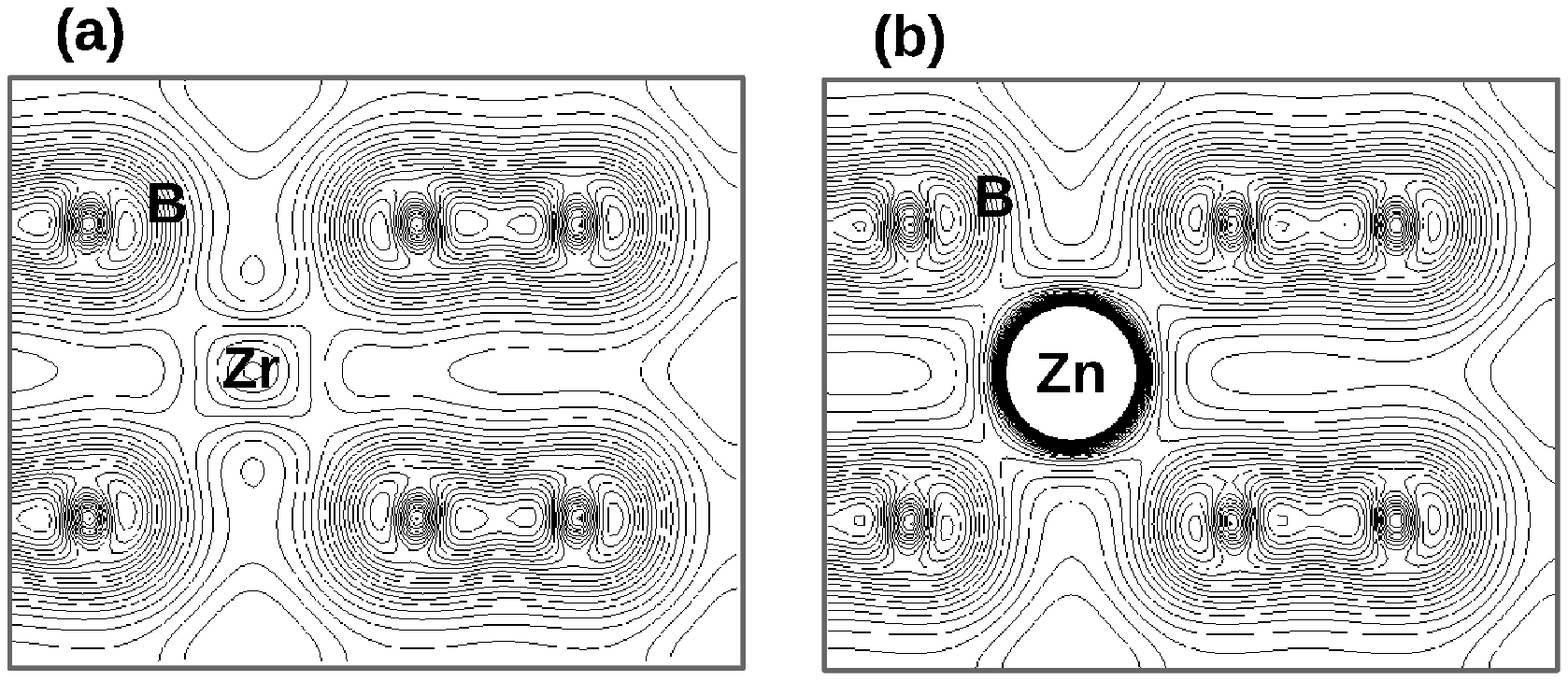}\hspace*{0em}
\vspace{2cm}
\caption{Charge density maps of 
(a) $\rm{Zr}$-substituted and (b) $\rm{Zn}$-substituted $\rm{MgB_2}$
projected onto the $\rm (100)$ plane. Calculations were made for a pressure of
$16.5$ GPa and $14.7$ GPa, respectively.}
\end{figure}


\section*{References}

\begin{thebibliography}{00}

\bibitem{Buz}
Buzea~C and Yamashita~T 2001 \SUST {\bf 14} R115

\bibitem{Rav}
Ravindran~P, Vajeeston~P, Vidya~R, Kjekshus~A and Fjellv{\r{a}}g~H 
2001 {\it Phys. Rev.} B {\bf 64} 224509

\bibitem{Mazin1}
Erwin~S~C and Mazin~I~I 2003 {\it Phys. Rev.} B {\bf 68} 132505

\bibitem{Ivan}
Ivanovskii~A~L 2003 {\it Fiz. Tverdogo Tela} {\bf 45} 1742
Engl. Trans.: 2003 {\it Phys. Solid State} {\bf 45} 1829

\bibitem{Cava}
Cava~R~J, Zandbergen~H~W and Inumaru~K 2003 {\it Physica} C {\bf 385} 8

\bibitem{Singh2}
Singh~P~P 2003 {\it Bull. Mater. Sci.} {\bf 26} 131

\bibitem{Bern}
Bernardini~F and Massidda~S 2006 {\it Europhys. Lett.} {\bf 76} 491

\bibitem{Castro}
Di~Castro~D, Ortolani~M, Cappelluti~E, Schade~U, Zhigadlo~N~D and Karpinski~J
2006 {\it Phys. Rev.} B {\bf 73}, 174509

\bibitem{Kortus2}
Kortus~J 2007 {\it Physica} C {\bf 456} 54

\bibitem{Kuzm}
Kuzmenko~A~B 2007 {\it Physica} C {\bf 456} 63

\bibitem{WLi}
Li~W~X, Zeng~R, Poh~C~K, Li~Y and Dou~S~X 2010 \JPCM {\bf 22} 135701

\bibitem{Ojha}
Ojha~N, Malik~V~K, Singla Rashmi, Bernhard~C and Varma~G~D
2010 \SUST {\bf 23} 045005

\bibitem{Singh1}
Singh~P~P 2002 {\it Physica} C {\bf 382} 381

\bibitem{Mor}
Moritomo~Y and Xu~Sh 2001 ArXive: cond-mat/0104568v1

\bibitem{Kaz}
Kazakov~S~M, Angst~M, Karpinski~J, Fita~I~M and Puzniak~R 2001 \SSC {\bf 119} 1

\bibitem{Tam}
Tampieri~A, Celotti~G., Sprio~S, Rinaldi~D, Barucca~G and Caciuffo~R
2002 \SSC {\bf 121} 497

\bibitem{Kir}
Singh~K, Mohan~R, Shelke~V, Gaur~N~K and Singh~R~K
2008 {\it Indian J. Pure \& Appl. Phys.} {\bf 46} 420

\bibitem{Kal}
Kalavathi~S and Divakar~C 2005 {\it Bull. Mater. Sci.} {\bf 28} 249

\bibitem{Nar}
Feng~Y, Zhao~Y, Yan~G, Pradhan~A~K, Zhou~L, Koshizuka~N and Murakami~M
2004 Critical Current Density, Flux Pinning and Microstructure in
$\rm{MgB_2}$ Superconductors
{\it High Temperature Superconductivity 1 Materials} ed~A~V~Narlikar
(Berlin, Heidelberg: Springer-Verlag) pp.~401-430

\bibitem{Sun}
Sun~Y, Yu~D, Liu~Z, He~J, Zhang~X, Tian~Y,
Xiang~J and Zheng~D 2007 Appl. Phys. Lett. {\bf 90} 052507

\bibitem{Shi}
Shi~L, Zhang~S and Zhang~H 2008 \SSC {\bf 147} 27
                                          
\bibitem{AM}
M'chirgui~A, Ben~Azzouz~F, Annabi~M, Zouaoui~M and Ben~Salem~M 2005
\SSC {\bf 133} 321

\bibitem{Mazin2}
Mazin~I~I and Antropov~V~P 2003 {\it Physica} C {\bf 385} 49

\bibitem{NN1}
Kristoffel~N, {\"{O}}rd~T and R{\"{a}}go~K 2003 {\it Europhys. Lett.} {\bf 61} 109

\bibitem{NN2}
Kristoffel~N, {\"{O}}rd~T and R{\"{a}}go~K 2003 {\it J. Supercond.} {\bf 16} 517

\bibitem{BHB}
Bussmann-Holder~A and Bianconi~A 2003 {\it Phys. Rev.} B {\bf 67} 132509

\bibitem{Kresse1}
Kresse~G and Hafner~J 1993 {\it Phys. Rev.} B {\bf 47} 558

\bibitem{Kresse2}
Kresse~G and Hafner~J 1994 {\it Phys. Rev.} B {\bf 49} 14251

\bibitem{Kresse3}
Kresse~G and Furthm{\"{u}}ller~J 1996 {\it Comput. Mater. Sci.} 6 , 15

\bibitem{Kresse4}
Kresse~G and Furthm{\"{u}}ller~J 1996 {\it Phys. Rev.} B {\bf 54} 11169

\bibitem{Blochl}
Bl{\"{o}}chl~P~E 1994 {\it Phys. Rev.} B {\bf 50} 17953

\bibitem{Kresse5}
Kresse~G and Joubert~D 1999 {\it Phys. Rev.} B {\bf 59} 1758
 
\bibitem{Perdew}
Perdew~J~P, Burke~K and Ernzerhof~M 1996 \PRL {\bf 77} 3865;
1997 \PRL {\bf 78} 1396


\bibitem{Monk}
Monkhorst~H~J and Pack~J~D 1976 {\it Phys. Rev.} B {\bf 13} 5188

\bibitem{Bader}
Bader~R~F~W 1990 Atoms in Molecules: A Quantum Theory
(Oxford:  Oxford University Press)

\bibitem{Henk}
http://theory.cm.utexas.edu/vtsttools/bader

\bibitem{Tang}
W.~Tang, E.~Sanville, G.~Henkelman 2009 \JPCM {\bf 21} 084204

\bibitem{Choi}
Choi~H~J, Roundy~D, Sun~H, Cohen~M~L and Louie~S~G
2002 {\it Nature} {\bf 418} 758

\bibitem{Mitr}
Mitrovi{\v{c}}~B 2004 \JPCM {\bf 16} 9013

\bibitem{EnvChem}
Barbalace K 1995-2011
{\it Periodic Table of Elements} (EnvironmentalChemistry.com)

\bibitem{Swit}
Switendick~A~C 1990
Electronic structure and charge density of zirconium diboride
{\it Technical Report} SAND--90-1474C; CONF-900870--7

\bibitem{An}
An~J~M and Pickett~W~E 2001 \PRL {\bf 86} 4366

\bibitem{Kortus1}
Kortus~J, Mazin~I~I, Belashchenko~K~D, Antropov~V~P and Boyer~L~L
2001 \PRL {\bf 86} 4656

\bibitem{Bel}
Belashchenko~K~D, van~Schilfgaarde~M and Antropov~V~P
2001 {\it Phys. Rev.} B {\bf 64} 092503

\bibitem{DeLaMora}
De~La~Mora~P, Castro~M and Taviz{\'{o}}n~G
2002 {\it J. Sol. Stat. Chem.} {\bf 169} 168

\bibitem{Moni1}
Monni~M, Ferdeghini~C, Putti~M, Manfrinetti~P, Palenzona~A, Affronte~M,
Postorino~P, Lavagnini~M, Sacchetti~A, Di Castro~D, Sacchetti~F, Petrillo~C and
Orecchini~A
2006 {\it Phys. Rev.} B {\bf 73} 214508

\bibitem{Cotton}
Kalvins~A~K 1985
A review of the inorganic and organometallic chemistry of zirconium
{\it Report No 85-124-K}

\bibitem{Shein1}
Shein~I~R and Ivanovskii~A~L 2002 {\it Fiz. Tverdogo Tela} {\bf 44} 1752
Engl. Trans.: 2002 {\it Phys. Solid State} {\bf 44} 1833

\bibitem{Rosner}
Rosner~H, An~J~M, Pickett~W~E and Drechsler~S-L 
2002 {\it Phys. Rev.} B {\bf 66} 024521

\bibitem{NN5A}
Kristoffel~N 2000 Structural distortions and oxygen local dynamic instabilities in superconducting perovskites
{\it Defects and Surface-Induced Effects in Advanced Perovskites}
({\it NATO Science Series 3. High Technology})
ed~G~Borstel, A~Krumins and D~Millers
(Dordrecht: Kluwer Scientific Publishers) pp.~113-124

\bibitem{NN5B}
Kristoffel~N and Klopov~M 1996 {\it Phys. Rev.} B {\bf 54} 9074

\bibitem{NN3}
Kristoffel~N~N 1984 {\it Czech. J. Phys.} B {\bf 34} 1253

\bibitem{NN4}
Kristoffel~N and Konsin~P 1988 {\it Phys. Status Solidi} (b) {\bf 149} 11

\bibitem{BH1}
Bussmann-Holder~A, Bilz~H and Benedek~G
1989 {\it Phys. Rev.} B {\bf 39} 9214

\bibitem{AP}
Pishtshev~A 2011 {\it Physica} B {\bf 406} 1586

\bibitem{Lee}
Lee~K~W and Pickett~W~E 2003 {\it Phys. Rev.} B {\bf 68} 085308

\bibitem{Kresin}
Bill~A, Morawitz~H and Kresin~V~Z 2003 {\it Phys. Rev.} B {\bf 68} 144519

\bibitem{Var}
Varshney~D, Azad~M~S and Singh~R~K
2004 \SUST {\bf 17} 1446\\
Varshney~D, Nagar~M, Bhatnagar~S and Varshney~M
2010 \SUST {\bf 23} 075016

\bibitem{Deng}
Deng~S, Simon~A and K{\"{o}}hler~J
2005 Pairing Mechanisms Viewed from Physics and Chemistry
{\it Superconductivity in Complex Systems}
({\it Structure and Bonding vol 114})
ed~K~A~M{\"{u}}ller and A~Bussmann-Holder 
(Berlin, Heidelberg: Springer-Verlag) pp.~103-141

\bibitem{Ghosez}
Ghosez~Ph and Veithen~M 2007 \JPCM {\bf 19} 096002

\end{thebibliography}
\end{document}